# From granular avalanches to fluid turbulences through oozing pastes
# A mesoscopic physically-based particle model

Annie Luciani

Institut National Polytechnique de Grenoble - ACROE
Grenoble France

## Abstract

In this paper, we describe how we can precisely produce complex and various dynamic morphological features such as structured and chaotic features which occur in sand pilings (piles, avalanches, internal collapses, arches), in flowing fluids (laminar flowing, Kelvin-Helmholtz and Von Karmann eddies), and in cohesive pastes (twist-and-turn oozing and packing) using only a single unified model, called "mesoscopic model". This model is a physically-based particle model whose behavior depends on only four simple, but easy to understand, physically-based parameters : elasticity, viscosity and their local areas of influence. It is fast to compute and easy to understand by non-physicist users.

*Keywords: Physically-based model, Particle modeling, granular material, fluids, pastes, matter state changes*

## 1. INTRODUCTION

Since the first use of physically-based modeling in Computer Animation and Simulation in the middle of the 80's [1][2], many works have been supported by physically-based particle models. But only few focus on changes in matter states. Most of these developed models for specific categories of natural phenomena. Several concern the simulation of fluid features [3][4][5][6][7][8]. Others concern the simulation of amorphous materials [9] and a very few concern the simulation of granular materials [10]. For the majority of these studies, the classical continuous formulation was chosen.

However, studying the changes in matter form is probably the best way to evaluate the level of genericity of a modeling method. It is also an efficient way of evaluating the perceptual quality of the motion by allowing the user to compare categories of commonly observed but nevertheless complex dynamic forms.

Many authors point out that the purpose of Computer Graphics in physically-based modeling is quite different from Physics. Particularly, two other qualities are required:

- Computer Graphics tools must be available to non-physicist users. The main challenge for these tools is to allow to the non-physicist user to model himself the phenomenon he wants, through a single method.

- Computer Graphics look for a high perceptual and cognitive quality of motion. Visual perception is probably the most adapted sensor to distinguish, evaluate, categorize, or compare fine and accurate complex dynamic morphological features: the consistency of a syrupy elastic paste; the difference between sand, sugar, and volcanic lava; and between steam or smoke particles.

If particle methods are the most generic methods, they are often seen as quite rough in the restitution of motion of accurate categories of complex phenomena.

Thus, our purpose is to design a single physically-based particle model capable of synthesizing fine dynamic characters of well-known complex phenomena, according to a clear specification of their main pertinent figures, such as:

- Granular materials and their well-shaped piles, avalanches and collapses; fluids and their stable and turbulent states; oozing pastes with their flowing twists and turns.

- states changes between them : from solid (rigid or deformable), to pure kinetic gas through granular collapsing, pastes, stable laminar and turbulent fluid.

## 2. RELATED WORKS

In 1989-1991, soon after the introduction of physical modeling in Computer Animation, four founding works were performed : Miller and Pearce [11], Terzopoulos and al. [12], Tonnesen [13], and Luciani and al. [14][15]. These studies differ on some aspects but they all point to a common issue: the need for a single unified model to simulate the different states of matter, and the need for a generic computer animation modeler.

At that time, these works explored only rough, simple figures of matter states and changes. Though these studies seem to be dated, they remain relevant because they are based on fundamental theories. They contain a large amount of potentialities for Computer Graphics and continue to play the role of pioneer on the non-solved question of genericity. It is surprising that they have not been more widely noted. It is probably the search for complex motions that has pushed computer graphics researchers to develop specific methods for categories of phenomena. The wide use of Navier-Stokes equation to simulate fluids is the most spectacular example.

The works referenced before were inspired by a common theory, investigated by Greenspan since 1973 [16][17]. But they adopt different solutions in adapting this theory. We analyze first the main ideas of these works. After, in the next paragraph (§3), we discuss their fundamentals in relation to basic concepts in Physics and we outline the basis of our approach.

The Greenspan approach assumes that all physical behavior emerges from the microscopic level and that it is possible to reproduce this behavior by considering only the laws which play at this level. Physical objects are described as set of interacting particles interacting one to another through a potential conservative law. This interaction function supports not only the simple molecular collision as in pure kinetic gas, but a cohesion interaction, like the well-known experimental Lennard-Jones interaction function.

Its general expression is then $F(D) = -\dfrac{a}{D^m} + \dfrac{b}{D^n}$, drawing three important features:

- It presents three zones: a pure attractive zone for large distances, a pure repulsive zone for very short distances, and an intermediate zone in which the combination of the attractive and repulsive effects causes the cohesion effect.
- It is mathematically expressed by a sum of two terms: the attractive term and the repulsive term.
- Each term is a non-linear function of the interparticle distance.

To simulate the granular or finer states (liquid or powder), the four works referenced before chose this Lennard-Jones interaction function between particles. For the representation of solid states, Tonnesen and Miller chose it also, while Terzopoulos and Luciani chose a visco-elastic interaction.

Miller adds a second pure viscous term between particles which depends on their relative velocities, introducing explicitly a damping function in the pairwise local interaction. Tonnesen introduces an ambient viscosity, which plays the role of a dissipation term with the exterior, making the system of particles as a whole dissipative. Terzopoulos introduces a model of dry friction with the other fixed objets of the scene. The results are simulations of only deformable solids, globular agglomerates, or curdled pastes and liquids.

In his simulation of fluid vortices, Greenspan himself introduces a damping velocity factor during the collision between particles and the cavity. Thus he simulates a dissipation during the interaction with fixed objects, being here the borders of the set of the particles. Thus, he created a wide variety of precise phenomena, but each was done under particular experimental conditions[17]: Cavity flow, turbulent and non turbulent vortices, liquid drop formation, fall and collision, fluid bubbles and jiggling gels, melting points ...

Aside from Greenspan-based approaches, Young and Mac Namara [18] have developed a different fundamental approach. Matter is represented by a set of hard disks coupled by inelastic collisions. Inelastic collisions are not conservative and involve a dissipation. Note that the previous repulsive interactions, with only a distance non-linearity, are not inelastic collisions. The regimes to look for are well identified: kinetic, shearing, clustering and collapsing. Inelastic collisions are represented by a restitution velocity factor r.

$$\vec{u}_1' = \vec{u}_1 - \frac{1}{2}(1+r)[\vec{k}\cdot(\vec{u}_1 - \vec{u}_2)]\vec{k}$$
$$\vec{u}_2' = \vec{u}_2 - \frac{1}{2}(1+r)[\vec{k}\cdot(\vec{u}_1 - \vec{u}_2)]\vec{k}$$

(u1,u2) and (u'1,u'2) : velocity vectors before and after the collisions - k : unit vector along the line of masses centers

By changing the restitution coefficient r, the authors prove that the main expected phenomena characterizing each matter state can occur spontaneously : free kinetic motion such as in pure gas, for r equal about 1; shearing state such as in pre-turbulent fluid for r between about 0.9 and 0.8; clustering state, such as in "gel" for r between about 0.8 and 0.6; and inelastic collapse such as in granular piling for r closed to 0.

They performed lengthy numerical simulations in which particles moved themselves according only to initial conditions.

## 3. FUNDAMENTALS

## 3.1 Discussion of the fundamentals

These approaches pose three fundamental questions : (i) what is the role of distance non-linearity on each term (mainly on the repulsive term with which one represents collisions, (ii) is the polynomial formulation convenient for the user, (iii) is the only potential interaction sufficient for all of these phenomena.

### 3.1.1 Non-linearities of the potential function

The use of a nonlinear potential collision interaction does not influence fundamentally the behavior of the system for the expected purposes. First, the system remains purely conservative. Second, this kind of non-linearity does not play a critical role in the behavior of fluids. In fact, it is usually used to simulate a matter-factor surrounding the particle. The elastic force between particles in contact varies non-linearly according to the deformation, simulating a variable compression. Perhaps, it would play a critical role in granular material behavior such as heterogeneousness in sand piles. But it has been experimentally proved in [19][20], that one can produce granular material effects with only the first order linear approximation of the potential collision interaction.

### 3.1.2 Lennard-Jones Polynomial formulation

The parameters of Lennard-Jones interaction forces are not easy to manipulate. But it is not because this function works at a microscopic level (as sometimes said), but because of the analytical formulation, the coefficients and the exponents of the polynomials act on the entire function. They do not permit to control explicitly and independently the critical parameters which regulate the shape and size of the zone of cohesion: repulsion and attraction slopes, rest distance of repulsion, distance thresholds between the attractive and repulsive part.

### 3.1.3 Viscosity and dissipation

Viscosity is crucial to particle modeling. Instead of elasticity which plays at a microscopic level, viscosity is not a primary microscopic component. It is one of the three basic transport phenomena besides the molecular diffusion described by the Fick law and the heat conduction described by the Fourier law. It is an emergent phenomenon. In a fluid flowing in a global convection motion, the molecular motion causes the molecules to be transferred along the convection velocity gradient, creating a mixing layer, with a transfer of momentum. Thus, at the macroscopic level, the fluid seems sheared by two equal and opposite forces in the direction of the convection motion. The proportionality factor between this force and the velocity gradient is called the viscosity coefficient.

Pure particle models propose that there is no need of viscosity at the microscopic level. But this leaves two scaling problems, related to the emergence of the viscosity effect. First, the number of particles exchanged in mixing layers must be sufficiently high to obtain a sufficient momentum exchange able to modify the macroscopic kinetic convection motion. To compensate the lack of the number of particles, Greenspan introduces a macroscopic viscous dissipation towards the exterior. Second, the simulation time rate must be consistent with the microscopic collisions, i.e. several times less than the one needed for macroscopic dynamics. The rate of the Greenspan's simulations is about 10KHz and the dynamical figures are observed each 500 frames, i.e., at 20 Hz. The scaling time factor is about 600.

Young and Mac Namara's experiments prove that inelastic collisions, i.e. with dissipation, are sufficient to regulate the different states of the matter. Their model is also easier to compute than Greenspan-based models. But the restitution coefficient will also be unwieldy to manipulate in complex heterogeneous scenes usually designed in computer graphics.

In conclusion, if these approaches are theoretically well founded, they are not the optimal set of primitive interaction functions. The formulation of the potential function is too complex and cryptic, and the formation of the dissipation function is too specific.

## 3.2 A mesoscopic single model

The Mac Namara and Greespan approaches, partially followed by Tonnesen and Miller, are "bottom to top" approaches or "from microscopic phenomena to macroscopic effects" approaches.

Our approach stands at a <u>mesoscopic</u> level, between a macroscopic and a microscopic level. It is the level sufficient enough to explain or reproduce the expected phenomena with a minimal set of comprehensive parameters.

As shown before, the fundamental idea is that matter evolves through its different states by balancing between potential and dissipation effects. As an interaction component, viscosity operates at this intermediate mesoscopic level. Placed between two particles, it defines them, not as molecules, but as a set of molecules, i.e. as a meso-component of matter, a kind of "matter parcel". When a potential component is placed between these meso-particles, it will represent not only a repulsive function but also the deformability of the particle after a contact. For a fluid, this meso-particle will represent a deformable parcel of fluid with a volume depending on an average number of molecules and an average free path. When two particles interact through a viscous interaction, we model thus the kinetic mesoscopic exchange between two pieces of mixing layers.

From these two sufficient components, the most simple model is obtained by the most simple expressions for each one, that is :

- for the potential component, a linear elastic function or a piecewise linear elastic function,
- for the viscous component, a linear viscous function or a piecewise linear viscous function.

$$F_{ij} = F_{ij}^e + F_{ij}^v$$
$$F_{ij}^e = \sum_k K_{ij}^k.(D_{ij} - D0_{ij}) \quad if \ DT1_k^e \leq D_{ij} \leq DT2_k^e, 0 \ if \ not$$
$$F_{ij}^v = \sum_k Z_{ij}^k.(D_{ij} - D0_{ij}) \quad if \ DT1_k^v \leq D_{ij} \leq DT2_k^v, 0 \ if \ not$$

By changing the parameters Kij, Zij, DT1, DT2, we obtain a family of potential functions representing a large variety of potential and viscous phenomena (Figure 1). In contrast to the fractional formulation, these functions are definite for D = 0.

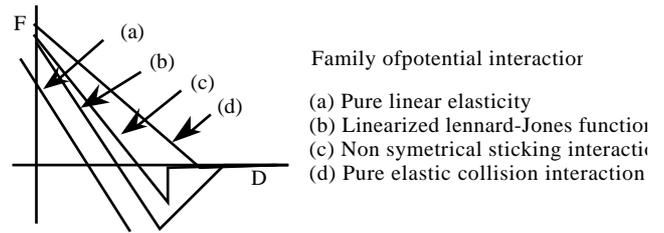

**Figure 1**: Family of interparticle potential interaction functions - (b) represents a linearized Lennard-Jones function.

Contrary to the parameters of the Lennard-Jones analytical formulations or the restitution velocity factor, the user has access explicitly to the critical parameters and these parameters have direct physical meaning for the user. Let us consider the simple case of elasticity and viscosity with a single threshold DTe and DTv, such as shown in the next figure (Figure 2).

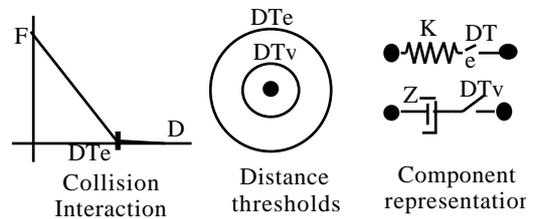

**Figure 2**: Simple thresholded viscous and elastic interactions

DTe (resp. DTv) represents the apparent elastic (resp. viscous) size of particles, which are exactly the sizes of the areas of influence of interaction functions. Thus, these thresholds have not only the role to decrease the complexity of the calculation (such as said usually in pure particle models), but above all, they have a physical signification. In fluids, this viscosity threshold gives the size of fluid parcels from which viscosity emerges and the elastic threshold helps to simulate pressure.

The slope of the function is explicitly the value of the elasticity (resp. viscosity), which regulates the apparent deformability of the meso-particles. It is easy to regulate independently the thresholds and the slopes, allowing a greater variety of functions than with the analytical formulation.

The dynamic behavior of a body is regulated by changing :

- the parameters of the potential function, in a similar but more convenient way than that of Tonnesen, Terzopoulos and Greenspan,
- by balancing explicitly the influences of the elastic and viscous terms through their elastic or viscous parameter, in a similar but simpler way than that of Miller or Young.

In Luciani [14], the genericity of this approach was presented at its beginning. We modeled rigidity, deep deformations and fractures. We showed fluid and powder behaviors such as those obtained by Miller and Tonnesen. Jimenez [15] modeled a plastic-like behavior with a linearized Lennard-Jones function (Figure 1-b) and an ambient viscosity term.

## 3.3 Our Requirements

With the preceding in mind, we formulate our hypothesis. Our purpose is to design a formalism, based on the philosophy of particles modeling, with the following differences:

- It must be fundamental and physically consistent, with a sufficient and necessary basis of elementary interparticle interactions. This formalism will be based on the two main and complementary interaction components : potential and dissipative, playing at the same level between meso-particles. In its main application, this formalism would allow the user to create complex heterogeneous Computer Graphics scenes.

- It must generate a large variety of well-specified physical phenomena with a small number of parameters according to their clear specification. We are challenged to achieve all the figures listed in the following (§4) for each of the five categories of matter states. We try to describe a larger variety of dynamic nonlinear state changes than those explored in the previous works.

- It should be convenient to manipulate by non-physicist users. Particularly, the parameters the user manipulates should be basic, distinct and simple to understand, and they should allow the user to create empirically his mental representation of each of them. We assume that it is the case of simple elasticity, simple viscosity and their space thresholds.

- By choosing the simplest formulation of interaction functions, (like piecewise linear functions rather than polynomial and fractional expressions), and with the concept of the mesoscopic model (running just-above the microscopic level, optimizing the computation rate and the number of particles), the computational cost would be kept low.

## 4. DYNAMIC FEATURES SPECIFICATIONS

To go further, we must characterize more precisely the main macroscopic features of the states of matter.

Matter is normally considered to exist in three states: gaseous, liquid and solid. As shown by Young and Mac Namara, for this kind of purpose on the state changes, we must also take into account characteristic differences between each of these states. Thus, we will consider two other states of matter: a granular and a gel state. Human perception and cognition are particularly well adapted to accurately and reliably analyze these dynamic effects of "sand", "oil", "water", "smoke", "creams", or "pastes".

## 4.1 Solid State

Solid bodies move and deform while preserving their individuality in time and space. This makes the modeling of solid regimes easy by linking particles with permanent interactions, such as a combination of visco-plasto-elastic interactions according to the matter of the solid.

## 4.2 Pure Gaseous Kinetic State

The pure kinetic gaseous state is also easy to model. By linking all particles with pure elastic collision interactions, we create a brownian motion in which the field of constraints is homogeneous, allowing the definition of macroscopic variables such as temperature and pressure.

## 4.3 Granular Material State

The behavior of granular materials, such as powders and sands, is not the same as a set of gas molecules. The main difference is that, the constraints inside the material are not homogeneous. The granular state is sometimes called "a fourth state of matter", between fluid and solid. The main features which allow to distinguish granular materials from fluids, solids, or gases are:

- Flows can evolve into piles.

- Granular piles exhibit specific geometrical effects. On an horizontal surface, the pile shape is a triangle in which the base equal angles characterize each type of material.

- The pile grows by chaotic surface avalanches.

- The constraints inside the pile are not homogeneous. For this, it is not possible to define a sort of granular temperature or granular pressure.

- When specific arrangements of internal constraints occur, auto-similar sub-piles appear, with constraint shearing between them. These sub-piles can translate as solids (they can also rotate), causing internal collapses.

## 4.4 Fluid State

There are three main macroscopic states in the fluid regime:
- Like in gases, constraints in fluids tend to be homogeneously shared. Under gravity, fluids spread on the recipient, as opposed to granular material piling.

- They can flow in a laminar fashion.

- They can be in a turbulent state with vortices and eddies.

The turbulent state exhibits characteristic figures such as Kelvin-Helmholtz and von Karman eddies. They occur with the emergence of a molecular mixing layer between fluid layers having different convection velocities. This mixing layer occurs neither in granular material nor in gases. In gases, particles can be completely mixed. When granular material is shaken, we observe a segregation effect: the biggest particles come up to the top of the granular agglomerate.

## 4.5 Gel and Pastes State

In the gel or paste regime, before amorphous agglomeration, there is a specific and very interesting behavior: the twist-and-turn oozing and folds shaping.

The typical behavior of gels, pastes and creams are the same as toothpaste when pushed from its tube. It twists during the flowing. When it meets a surface, the twist-and-turn motion generates folds which evolve specifically during time: the folds regress slowly and the paste tends to sink, pack and spread out onto the surface, according to its fluidity and compressibility, being like either a foam, or a cream or a paste.

# 5. A SINGLE MODEL FOR FIVE STATES

As previously mentioned, an association between pure thresholded elastic interaction and pure thresholded viscous interaction would be theoretically sufficient to obtain the five categories of phenomena. From this, we define then the generic qualitative structure of the model, only composed of these two interactions in parallel as drawn in figure 3-IP5, in which:

K is the interparticle elasticity parameter,
Z is the interparticle viscosity parameter,
De is the elasticity distance threshold defining the distance locality in which the elastic interaction plays.
Dv is the viscosity distance threshold defining the distance locality in which the viscous interaction plays.
The following figures show the different results of experiments on the thresholded viscous-elastic model.

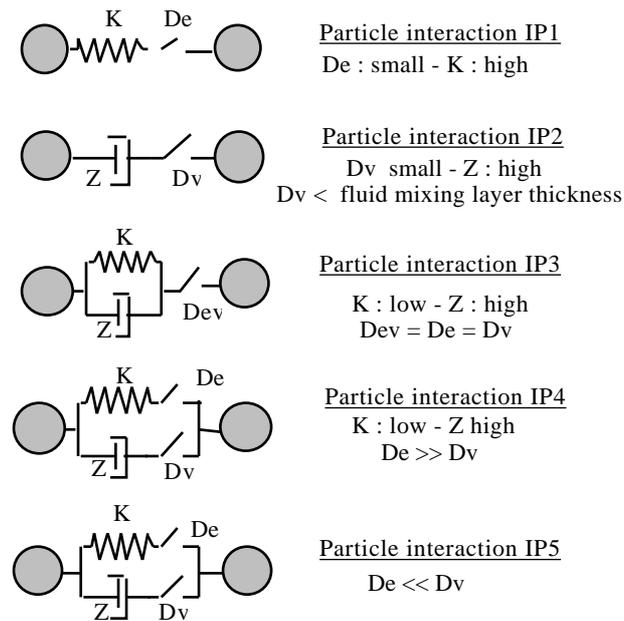

**Figure 3**: Different parametric versions of the single model simulating the five matter states

### 5.1.1 Pure Kinetic Gas behavior

Pure kinetic gas behavior is simply modeled with IP1 interaction. Figure 4 shows snapshots of a simulation in which two laminar jets collide and evolves in a pure chaotic motion.

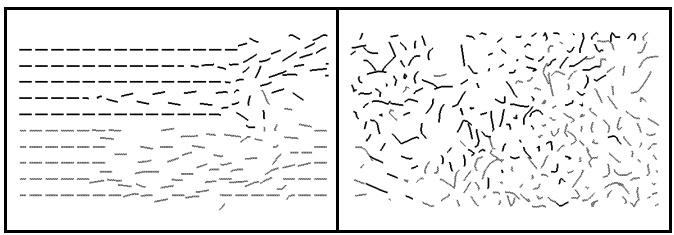

**Figure 4**: Snapshots of a simulation of pure kinetic gas.

### 5.1.2 Granular material behavior

The main figures of granular material are obtained with the particle IP1 interaction (Figure 3). The pertinent parameters are the value of the elasticity between particles and the ground friction. Ground friction is modeled with IP1 interaction in which the distance threshold defining the coarseness of the ground particles is chosen to be small in comparison to the particles themselves. There is no dissipation between particles. We place only an ambient viscous non thresholded interaction which plays equally on all the particles.

The simulation snapshots 5(a), 5(b), 5(c), 5(d) show the main features of granular material dynamic effects. (Note[1])

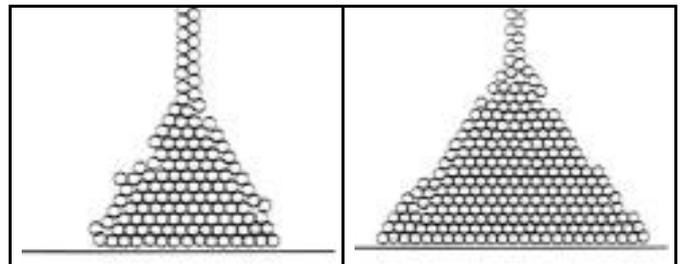

5(a) Characteristic pilings

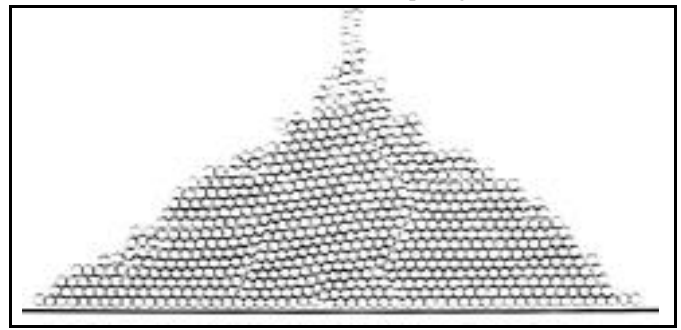

5(b) Heterogeneousness of inner constraints

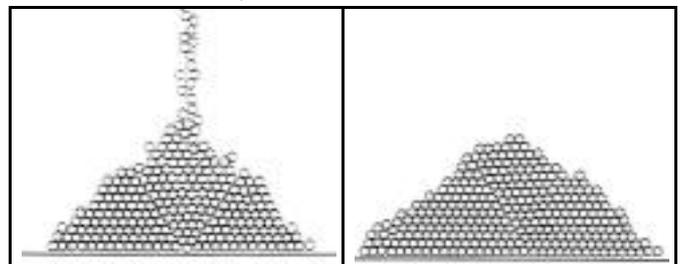

5(c) auto-similar sub-piles with internal collapses

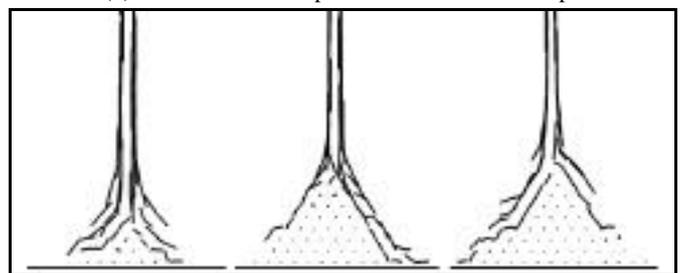

5(d) Surface avalanches

**Figure 5**: Snapshots of granular material simulations

---

[1] The quality of the snapshots can be altered by the PDF compaction process.

### 5.1.3 From pure dry to moist granular material behavior and amorphous paste

Decreasing the value of the elasticity and increasing the value of the viscosity, the state changes from dry piles to amorphous material through more or less moist piles. Notice that the clearly cut lines of linear shearing disappear progressively.
Figure 6 shows snapshots of four simulations of different granular states, from dry to moist.

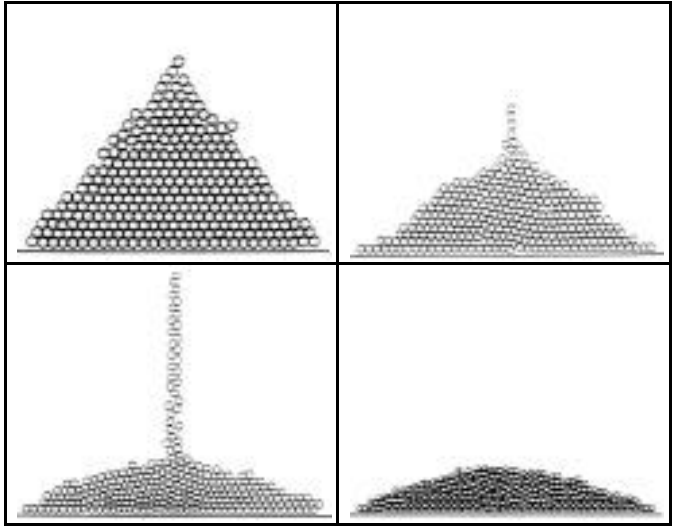

**Figure 6**: Snapshots of 4 simulations of different granular states - from dry state (Up left) to moist state (down right).

### 5.1.4 Turbulent fluid behavior

We obtain the main figures of fluid turbulences using the particle interactions IP2, IP3 and IP4 (Figure 3).

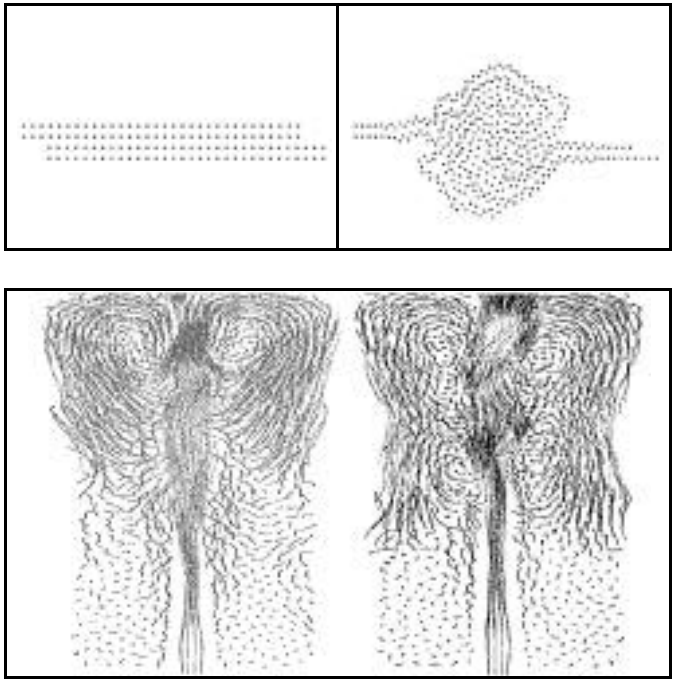

**Figure 7**: Snapshots of simulation of turbulent fluids

The purely thresholded viscous IP2 interaction is sufficient to obtain Kelvin-Helmholtz eddies. To add von Karmann paths figures, in which the inner fluid pressure plays a role, we simulate the pressure with a thresholded elastic interaction (IP4 model) with a threshold greater than the threshold of the viscous interaction and with a low elasticity: K low - Z high - De high - Dv low. This means that the elastic confinement is greater than the viscous confinement. Therefore, each particle can interact with more particles from the elastic interaction than from the viscous interaction : particles can be in a collision interaction without being in a viscous interaction.

Figure 7-up shows the birth of Kelvin Helmholtz eddy from two laminar opposite jets. Figure 7-low shows turbulences and their propagation in a interacting jet -medium object.

### 5.1.5 From fluid spreading to granular piling

If the viscous and elastic thresholds are equal, as obtained with the IP3 interaction, then the spatial influence of the viscosity and the elasticity are the same. The particles which are confined in the collision area, are simultaneously linked by viscous and elastic interactions. Then when we increase the elasticity, as shown in the parameter-behavior map drawn in figure 8, the behavior evolves from a more or less compressible viscous fluid behavior to granular material behavior.

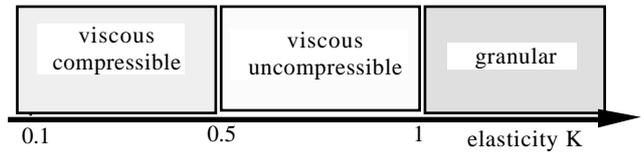

**Figure 8**: Visco-elastic thresholded interaction IP3 - Behavior according to the value of the elasticity

Figure 9 shows snapshots of four simulations experimenting the transient between fluid spreading and granular piling.

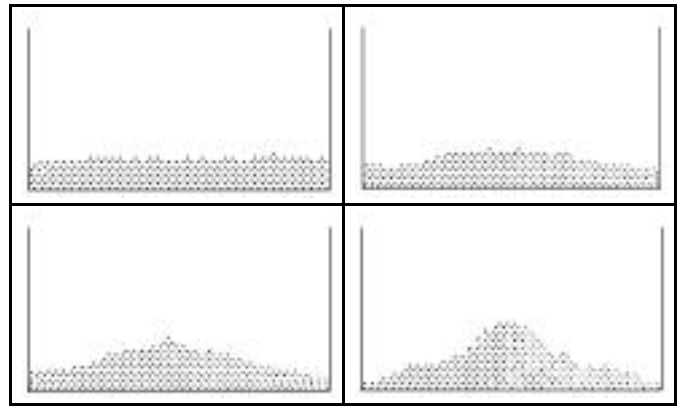

**Figure 9**: Last images of 4 simulations: from fluids to granular materials - Up left : Fluid spreading after flowing - Down right : Sand pile after flowing

### 5.1.6 Structured pastes behavior

If the elastic threshold is less than the viscous threshold, all the particles being in elastic contact will be also in viscous interaction and particles will be in viscous interaction

International Conference Graphicon 2000, Moscow, Russia, http://www.graphicon.ru/

without being in elastic contact. Then, the number of elastic collisions on one particle will be reduced while the viscous cohesion, evaluated as the number of particles viscously linked to this particle, will be increased (Figure 8). The viscous behavior is then predominant in a certain area around the particle. This constitutes the IP5 interaction producing various paste and cream behaviors. Figure 10 compares the local viscous and elastic confinements in both cases, fluids and pastes. In the area between Dv and De, the behavior is elastic for fluids and viscous for pastes. In close proximity to each particle, the predominant behavior is viscous for fluids and elastic for pastes.

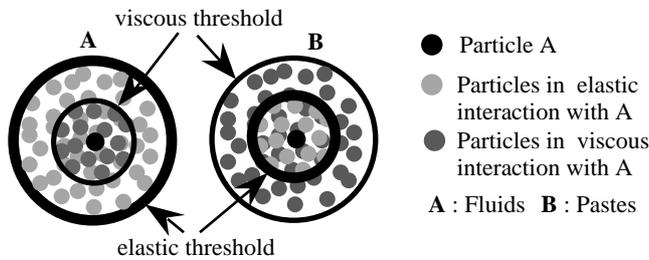

**Figure 10**: Elasticity and Viscosity influences in fluids and pastes. On the left, the fluid's model - On the right, the paste's model.

By varying the ratio between Dv and De, the behavior changes continuously from fluids to different pastes. Figure 11 shows 12 frames of a simulation of a paste having high elasticity and high viscosity and for which the threshold ratio is Rs = Dv/De = 4. The shown frames are every 4 seconds. Note the twist-and-turn oozing, the piling with swirling folds, a few with horizontal spreading and a few with vertical packing.

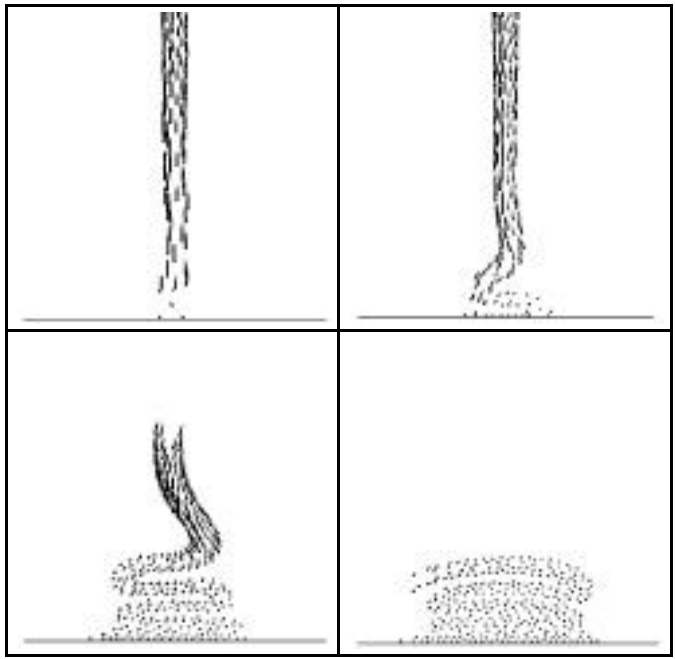

**Figure 11**: Snapshots of a paste simulation

### 5.1.7 From Granular material to creamy consistency - Influence of the viscosity threshold

As shown in the figure 12, with the values of elasticity and viscosity similar to those making the granular features, we increase the threshold ratio Rs = Dv/Dk from 1 to 4, the material changes from a granular behavior to a creamy consistency. From any value of this ratio, the increase of viscosity does not make a critical change, i.e. the increase of the viscous interaction threshold makes the creamy consistency more possible than an increase of the value of viscosity.

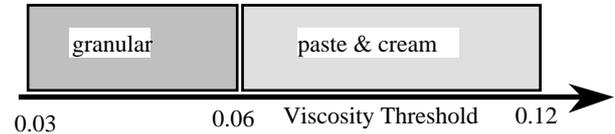

**Figure 12**: Change of consistency in function of viscosity threshold

### 5.1.8 General behavior parametric graph

Figure 13 shows the behavior of the model in the plane (K, Rs = Dv/De); the value of the viscosity remaining medium :

- Zone A3 : for high values of elasticity in the collisions interactions, we observe the emergence of granular behavior.

- Zones A1 and A2 : At low elasticity and low thresholds ratio, we obtain fluids behavior, a quite compressible behavior in A1 and quite uncompressible in A2.

- Zones B1 and B2 : When we increase the threshold ratio Rs (increasing the viscous threshold), we obtain a more creamy consistency : compressible in B1 such as in foams and more uncompressible in B2 such as doughy pastes.

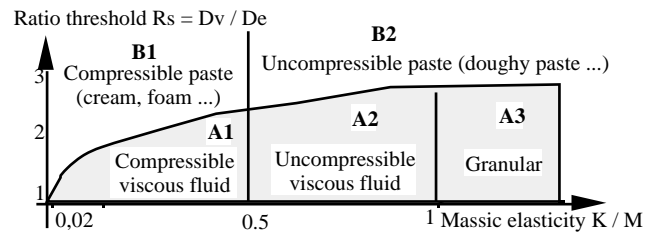

**Figure 13**: Pasty consistency in the plane (K, Rs).

## 6. CONCLUSION

Theoretically, the concept of mesoscopic model positions the fundamental Particle Modeling Theory on the field of efficient computer modeling and simulation for non-physicist uses.

Experimentally, the experiments described in this studies show that it is possible to simulate fine categories of complex dynamical effects with physically-based particle approaches: all the chaotic and structured figures of granular materials, main figures of fluids flowing and turbulences, various behaviors of pastes oozing and spreading.

The behavior parametric graphs, qualifying the observed behaviors of the model according to a previous well-defined specification of the dynamical effects and quantitative parameters values, are tools for guiding the user to travel in the space of the parameters which regulate the states and their changes.

The qualitative guide for the physical and perceptual analysis is centered on the balance with what it can be called a viscosiy density and an elasticity density. According to the kind of dominating force, we explore a large range of dynamical figures of matter's behavior.

Because of the optimization of the number of particles due to the mesoscopic approach (particularly with the use of an interparticle viscosity) and the optimization of the expression of basic functions, the calculation time is less than this obtained with classical methods. The simulations concern about 300 particles for the 2D snapshots shown in the paper. The computations are always in 3D. For equivalent 3D scenes, the number of particles is about 1000 particles. The simulation rate is 1050Hz. Without any classical optimizations processes (such as the limitation of the calculations inside an adapted box), the simulation time on a Silicon Graphics O2-150 MHz is about 30 seconds by image frame with 25 image frames per second.

To achieve these studies, four kind of studies were performed :

- a theoretical analysis to converge to a single model, which analyzes the physical influence of physical elements
- a list of phenomena specifications, in which we pose clearly the critical figures to be achieved
- an experimental protocol, composed of a common experimental background for all the experiments.
- a precise experimentation methodology, to explore efficiently all the parameters space.

## 7. ACKNOWLEDGMENTS

These works have been received the financial support of the Ministère de la Culture of France.## 8. REFERENCES

[1] Terzopoulos, D. Elastically Deformable Models. Computer Graphics, Vol. 21, N°4, 205-214, 1987

[2] Luciani, A. and Cadoz C. Utilisation de modèles mécaniques et géométriques pour la synthèse sonore et le contrôle d'images animées, 2$^{ième}$ Colloque image, CESTA, Nice, France, 1986.

[3] Stam, J. A General Animation Framework for Gaseous Phenomena. ERCIM Research Report, R047, January 1997

[4] Stam, J. Stable Fluids. Proc. of SIGGRAPH'99, 121-128, 1999.

[5] Chiba, N. , Sanakanishi, S., Yokoyama, K., Ootawara, I.. Visual Simulation of Water Currents Using a Particle-based Behavioral Model. The Journal of Visualization and Computer Animation, Vol. 6, 155-171, 1995.

[6] Foster, N., Metaxas, D. Realistic Animation of Liquids. Graphical Models and Image Processing. 58(5): 471-483, 1996.

[7] Foster, N., Metaxas, D. Modeling the Motion of a Hot Turbulent Gas. Proc. of SIGGRAPH'97, 181-188, 1997.

[8] Witting, P. Computational Fluid Dynamics in a Traditional Animation Environment. Proc. of SIGGRAPH'99, 129-136, 1999.

[9] Desbrun, M., Gascuel, M-P. Animating soft substances with implicit surfaces. Proc. of SIGGRAPH'95, 287-290, 1995

[10] Xin Li, J., Moshell, M. Modeling Soil : Realtime Dynamic Model for Soil Slippage and Manipulation. Proc of SIGGRAPG'96, 1996.

[11] Miller, G., and Pearce, A. Globular Dynamics : A Connected particle System for Animating Viscous Fluids. Computer & Graphics Vol. 13, N°3,305-309, 1989.

[12] Terzopoulos, D., Platt, J. and K. Fleisher. Heating and melting deformable models : from goop to glop. Graphics Interface'89,  219-226, 1989.

[13] Tonnesen. D. Modeling liquids and solids using  thermal particles. Graphics Interface'91, 255-262, 1991.

[14] Luciani, A. Jimenez, S., Cadoz, C., Florens, JL., Raoult, O. Computational Physics : A Modeler-Simulator for Animated Physical Objects, Proceeding of Eurographics Conference'91, Elsevier Ed., 1991

[15] Luciani, A., Jimenez, S., Raoult, O., Cadoz C., Florens, JL. An unified view of multitude behaviour, flexibility, plasticity and fractures: balls, bubbles and agglomerates. Modeling in Computer Graphics, Springer Verlag (Ed.), 54-74, 1991.

[16] Greenspan, D. Discrete Models. Reading in Applied Mathematics. Addison-Wesley, 1973

[17] Greenspan, D. Particle Modeling. Birkhauser Ed., 1997.

[18] Young, W. and McNamara, S. Dynamics of a freely evolving two dimensional granular medium. Physical review, 53(5), 1996.

[19] Luciani, A., Habibi, A. and Manzotti., E. A multi-scale physical model of granular materials. Proccedings of Graphics Interface'95, 1995.

[20] Luding, S., Duran, J., Clément, E., Rajchenbach, J. Simulations of Dense Granular Flow : Dynamic Arches and Spin Organization. J. Physics I, France 6, 823-836, 1996.

[21] Luciani, A., Habibi, A., Vapillon, A., and Duroc, Y. A physical model of turbulent fluids. Computer Animation and Simulation, Eurographics'95, pages 16-29, 1995.## About the author

Annie Luciani,

Institut National Polytechnique de Grenoble,

46 av. Félix Viallet,

38031 Grenoble Cedex, France

E-mail: **Annie.Luciani@imag.fr**International Conference Graphicon 2000, Moscow, Russia, http://www.graphicon.ru/